\documentclass{modified}

\begin{document}

\title{LINE SHAPE ANALYSIS OF LINEAR X RAY MAGNETIC\\
SCATTERING COPT THIN FILMS
}

\author{\footnotesize E.V.R.CHAN}

\address{ University of Washington,Box 351560\\
Seattle, Washington, 98195-2420, United States.\\
evr@u.washington.edu}

\maketitle

\begin{abstract}
Data analysis of the CCD files from x ray magnetic 
resonance scattering linearly polarized in transmission 
geometry produces information about the radial and
azimuthal intensities.  In a series of measurements 
of increasing photon energies trends in data are
analyzed. 
\end{abstract}

\section{Introduction}

Magnetic thin film systems and multilayer systems have
been studies very actively because of their magnetic
properties and possible application for practical
devices, such as magnetic recording media technologies.

\section{Experiment}

Samples were grown on smooth, low-stress, 160 nm. thick
SiNx membranes by magnetron sputtering; they all had
20 nm. thick Pt buffer layers and 3 nm. thick Pt caps. 
Between the buffer layer and the cap, the samples had 
50 repeating units of a 0.4 nm. thick Cobalt layer and
a 0.7 nm. thick Pt layer.  Experiments used linearly 
polarized x rays from the Advanced Light Source at 
Lawrence Berkeley Laboratory (supported by USDOE),
the thirteenth harmonic of the beamline 9, undulator  
gap of 54 mm. near the resonant Cobalt L edge.  
To achieve transverse coherence, the raw 
undulator beam was passed through a 35 micron diameter 
pinhole before being scattered in transmission by 
the sample.  The distance from the sample to the CCD 
is 118 cm.  The resonant magnetic scattering was
collected by the Princeton soft x-ray CCD camera 
1024 X 1024 pixels in an area one inch by one 
inch.  The intensity of the raw undulator beam
was 2 X 10 \verb_**_ 14 photons/sec., the intensity of the 
coherent beam was 2 X 10 \verb_**_  12 photons/sec., and
the intensity of the scattered beam, was 2 X 10 \verb_**_  7 
photons/sec.  Each speckle pattern was measured 
for 30 to 100 seconds, so the total number of
photons in each CCD image 1024 X 1024 pixels 
is about 10 \verb_**_  9.  The speckle patterns may
be used to reconstruct the magnetic domain 
structure of the sampl$e^{1-3}$; 
this is but one of a general class of the old 
inverse or phase retrieval problem$s^4$.

\begin{figure}[th]
\centerline{\psfig{file=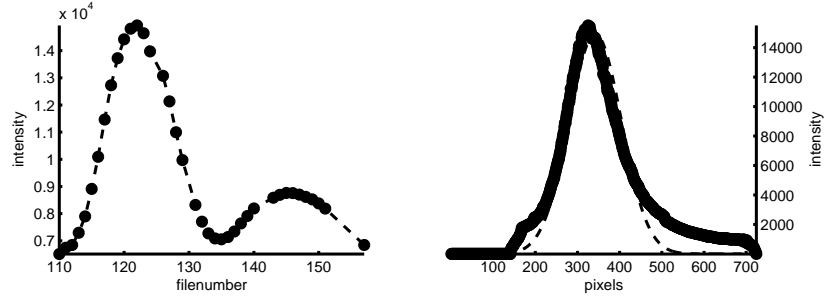,width=11cm}}
\vspace*{8pt}
\caption{(left) Average intensity versus filenumber 
for the series of images.  
 (right) Azimuthally averaged
radial intensity of image file 122.} 
\end{figure}

\section{Data Analysis}

The data appears as a series of CCD image files 
726Axxx.SPE (of increasing photon energy) that
is read into the data processing computer program.
The file numbers can be converted to photon energy
since the difference between the two maxima files 122
(photon energy 779.2 eV.) and 145 is 15.8 eV.
Figure~1 (left) shows the magnetic resonance has
two peaks.  Each piece of data has an image file
associated and that file has a 1024 X 1024 matrix 
containing intensity values ( if plotted in 3D 
it has the shape of a centered crown ). 
The Princeton CCD camera 
file is read into the freely available Matlab 
data analysis program by the following code 
fragment:  
\begin{verbatim}
%auto-ignore 
fid=fopen('nameOfFile.SPE','r');
header=fread(fid,2050,'uinit16'); %half of 4100 bytes
ImMat=fread(fid,1024*1024,'uint16');
Z=reshape(ImMat,1024,1024);
fclose(fid);Z=double(Z);
[X,Y]=meshgrid(1:1024,1:1024);
mesh(X,Y,Z); %display 3D plot
axis square; axis tight; view(90,90);
print -djpeg99 nameOfFile.jpg %highest resolution saved
%send email  evr@u.washington.edu
\end{verbatim}  
Figure~1(right) shows the variation of the intensity in 
file 122 in the radial direction that has been 
azimuthally averaged (consider slicing through 
the center out past the edge of the crown).  The
data analysis calculates the radius of each pixel, 
places it into appropriate bins and finds the 
average intensity; since each bin is only one 
pixel wide the bin number is the radius rounded 
to integer.  The profile was fitted to a gaussian
using non-linear Levenberg-Marquardt least squares.  
The difference between the maximum and minimum 
calculated values is the height, the distance of 
the maximum from the origin is called the peak 
center, and the FWHM (full width at half maximum) 
is called the width.  More about the variation
of the height, width and center with filenumber
will appear at the end.

\section{Results and Discussion}

The series of files 726Axxx.SPE are analysed for normalized
cross-correlatio$n^5$ using 2D matrices of the image 
files from which the average pixel values have been subtracted.  
The sum of the product of pixel values from two images divided
by the square root of the product of the sum of the pixel values 
squared of each image (makes it normalized) is gamma 
(normalized cross-correlation) of the two images.  If 
gamma is 1 they are correlated, if zero uncorrelated 
and if -1 anti-correlated.  The normalized 
cross-correlation function, gamma, is related to the 
coherence functio$n^6$. Table 1 shows the values of 
gamma for four of the files selected as file 
117(middle of first peak), file 122 (top of first peak), 
file 133 (minimum between the two peaks) and file 145 
(top of second peak).

\begin{table}[pt]
\tbl{  Normalized      Cross-Correlation}
{\begin{tabular}{@{}ccccc@{}}\toprule
   &  117 & 122 & 133 & 145 \\
\colrule
117 & 1.0 & 0.8 & 
-0.8 & -0.8 \\
 122 & 0.8 & 1.0 & 
  -0.9 & -0.9 \\
133 & -0.8 & -0.9 & 1.0 & 
\hphantom{0}0.9 \\
145 & -0.8 & -0.9 & 0.9 & \hphantom{0}1.0\\ \botrule
\end{tabular}}
\end{table}

The CCD images were processed so as to remove the burns,
remove anomalous charge scattering, remove the blocker arm, 
centered in the image and the central disk darkened to 
remove the burns in that area also.  The bright spots on 
the image are fixed burns in the CCD camera that cause a 
few pixels to be unusable; these are the ones that 
appear very bright. At the end of a series of SPE image
files a CCD burns only image was taken and a burns and 
anomalous charge scattering image was taken.  The difference 
of these two images was used to subract off the anomalous 
charge scattering pisels from each image file pixels 
(provided they are above the minimum background value 
of the image).  After the anomalous charge scattering is 
removed, the burns (or hot spots) are removed.  The hot 
spots are masked off, and after rotating around the 
middle of the 1024 X 1024 image, new background replaces 
the hot spots.  Next, the circle of maximum intensity is 
found by having an imaginary turtle going out from beyond 
 the edge of the blocker disc in rays every 1 degree 
finding maximum intensity. The intensities are sorted, 
the lowest 20 points out of 360 dropped and a least 
squares fit of the 340 points is done yielding the circle 
of maximum intensity.  The approximate edges of the 
blocker arm (about 7 degrees) are located by finding 
those points on the circle of maximum intensity where 
the second derivative changes sign.  A mask is created, 
a rotation picks up a patch, and the pixels of the 
blocker arm are replaced.  Missing background is fille 
in on the edges to make 1024 X 1024.  All pixels further 
out from the center, beyond the circle of maximum 
intensity were used to find the centroid and then the 
whole pattern was relocated to the center of the image 
matrix.

\begin{figure}[th]
\centerline{\psfig{file=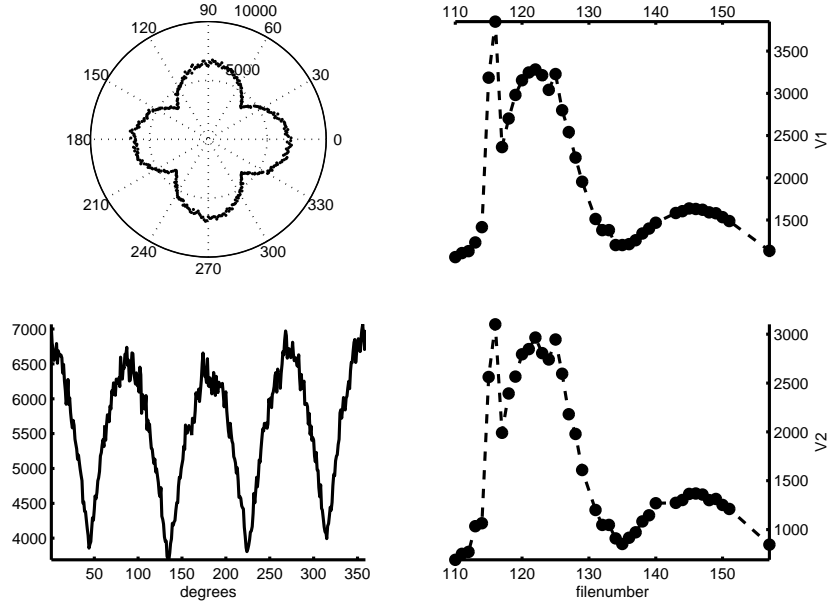,width=11cm}}
\vspace*{8pt}
\caption{(upper left)Polar plot of average azimuthal angular
intensity looking down the z axis at file 122 (lower left)
 azimuthal angular intensity of file 122  (upper right )
 V1 versus filenumber where V1 is the difference between the 
first maximum azimuthal radially averaged intensity and the 
lowest minimum (lower right) V2 versus filenumber where V2
is the difference between the second maximum azimuthal
radially averaged intensity and the lowest minimum}
\end{figure}

Due to the use of coherent light from synchrotron radiation, 
the scattering patterns produced are highly speckled.  
Each speckle is the sum of light scattered from all 
the illuminated magnetic domains.  So small changes 
in the microscopic orientation of the magnetic domains 
can have a large effect on the speckle pattern.  The bulk 
magnetization of Cobalt usually is in the x-y plane 
but when there are only a few Cobalt atoms in the layer 
the magnetication becomes perpendicular to the plane of 
the layers.  The observed intensity is related to the 
charge density, polarization, photon energy, magnetization, 
atomic scattering factor and scattering geometry; there are 
terms due to magnetic resonance and electronic structur$e^7$.

The azimuthal variation of intensity is calculated by 
looking at the angle from the center for each pixel, 
classifying it as belonging to bins (1 to 360) and 
finding the average of each bin; this is plotted 
versus the bin number with each bin being one degree 
wide.  In Fig.~2 the difference between the maxima and the 
lowest minimum is V1, V2.  The azimuthal variation of the 
intensity is described by an equation for magneto-crystalline 
anisotropy energy that includes a term proportional to 
the sine of twice the azimuthal angle square$d^8$.
The amplitude has real and imaginary charge and magnetic 
anomalous scattering factors which are tensors in  the 
general cas$e^9$.

The right side of Fig.~2 pretty much follows the shape of 
average intensity versus file number in Fig.~1(left) 
and illustrates the variation of the magnetic anisotropy 
difference (maximum minus lowest minimum) with photon 
energy(increases with file number).  The average of 
the value of first maximum and second maximum would be 
a reasonable estimate of the variable.  The reason the 
magnetic anistropy has these variations is because 
the atomic factor is a tenso$r^{10}$.

\begin{figure}[th]
\centerline{\psfig{file=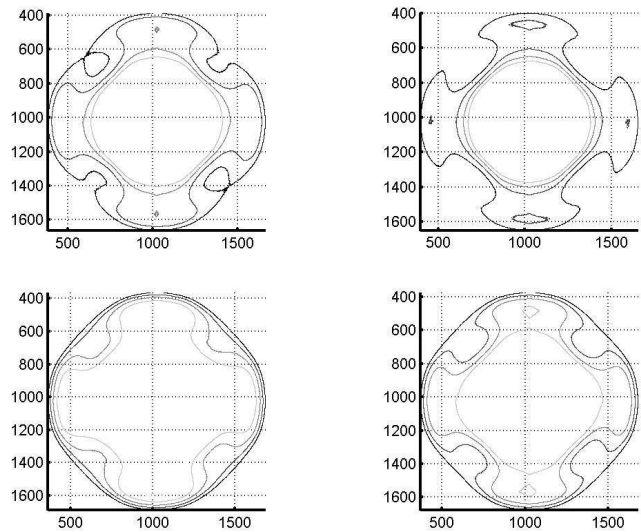,width=11cm}}
\vspace*{8pt}
\caption{(upper left)contour plot of autocorrelation 
file 117 halfway up first peak, (lower left)
contour plot of autocorrelation file 133 minimum
between peaks,  (upper right )contour plot of autocorrelation
file 122 tope of first peak, (lower right) 
contour plot of autocorrelation file 145 top of second peak}
\end{figure}

The autocorrelation has a maximum at the origin.  You 
could think of it as the convolution of the complex 
conjugate of f(-x,-y) and f(x,y).  If f(x,y) has a 
Fourier transform F(s,r), then its autocorrelation 
function has the transform absolute value squared of 
F(s,r) and has no phase information(Wiener-Khinchin 
theorem).  In the data analysis, the autocorrelation 
is evaluated by taking the Fourier transform of the 
reverse complex conjugate of the image and the Fourier 
transform  of the image, then taking the real part 
of the inverse Fourier transform of the product of 
these two Fourier transforms.  The three dimensional 
plot of the autocorrelation looks like a mountain with 
a narrow spike in the middle.  The threshholded (25-30\%) 
contours(5) in the x-y plane of the different files 
show more variation and different symmetries.  
Figure~3(upper left) File 117, halfway up the first 
peak, looks like it has four 2-fold axes.  
Figure~3(upper right) File 122, at the top of the 
first peak, may have one 4-fold and two 2-fold axes.  
Figure~3(lower left) File 133, the minimum between 
the peaks, appears to have one 4-fold axis and two 
2-fold axes.  Figure~3(lower right) File 145, top 
of the second peak, seems to have four 2-fold axes.

\begin{figure}[th]
\centerline{\psfig{file=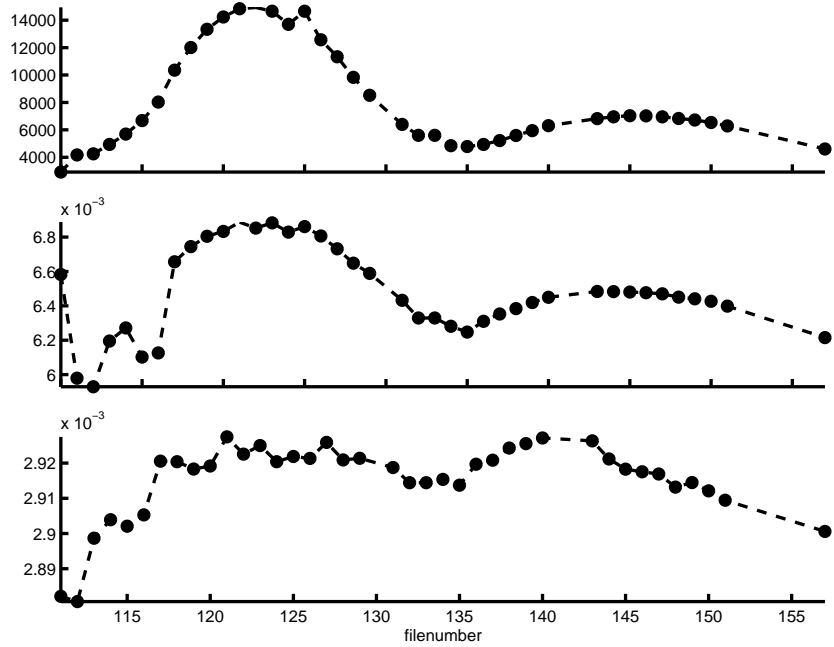,width=11cm}}
\vspace*{8pt}
\caption{(top) Peak height from the fitted gaussian versus
filenumber (middle) Inverse of the full width at half 
maximum versus filenumber (bottom) Inverse of center of
peak versus filenumber}
\end{figure}

In Fig.~4(top), the height of the peak verus file number is
plotted and is in agreement with Fig.~1(left).  The correlation 
length is related to the inverse of the width.  In 
Fig.~4(bottom), the inverse of the peak center is plotted 
versus file number; the spacing of the magnetic domains is 
inverse to the peak center.  In all cases the curves 
follow Fig.~1(left) more or less; the last one doesn't
follow the second peak very well.

\section{Conclusions}

It is possible to see azimuthal intensity variation 
through data anaylsis of CCD images.  Because of the odd 
behavior of the second peak and the differences in the 
autocorrelation functions(reflecting differences in 
electronic configurations) of the first and second peaks, 
there appears to be a difference in the nature of the 
first and second peaks.


\end{document}